**IET Communications**

ORIGINAL RESEARCH

# Over-the-air equalization with reconfigurable intelligent surfaces

Emre Arslan[1] | Ibrahim Yildirim[1,2] | Fatih Kilinc[1] | Ertugrul Basar[1]

[1]Department of Electrical and Electronics Engineering, Koç University, Sariyer, Istanbul, Turkey

[2]Faculty of Electrical and Electronics Engineering, Istanbul Technical University, Istanbul, Turkey

**Correspondence**
Emre Arslan, Department of Electrical and Electronics Engineering, Koç University, Sariyer 34450, Istanbul, Turkey.
Email: earslan18@ku.edu.tr

**Abstract**

Reconfigurable intelligent surface (RIS)-empowered communications is on the rise and is a promising technology envisioned to aid in 6G and beyond wireless communication networks. RISs can manipulate impinging waves through their electromagnetic elements enabling some sort of control over the wireless channel. The potential of RIS technology is explored to perform a sort of virtual equalization over-the-air for frequency-selective channels, whereas equalization is generally conducted at either the transmitter or receiver in conventional communication systems. Specifically, using an RIS, the frequency-selective channel from the transmitter to the RIS is transformed to a frequency-flat channel through elimination of inter-symbol interference (ISI) components at the receiver. ISI is eliminated by adjusting the phases of impinging signals particularly to maximize the incoming signal of the strongest tap. First, a general end-to-end system model is provided and a continuous to discrete-time signal model is presented. Subsequently, a probabilistic analysis for elimination of ISI terms is conducted and reinforced with computer simulations. Furthermore, a theoretical error probability analysis is performed along with computer simulations. It is analysed and demonstrated that conventional RIS phase alignment methods can successfully eliminate ISI, and the RIS-aided communication channel can be converted from frequency-selective to frequency-flat.

## 1 | INTRODUCTION

Commercial fifth generation (5G) wireless network standards have successfully been developed and countries are striving to adapt to 5G as means to not trail behind. Meanwhile, the race for the next generation wireless has already begun as academia and industry advance towards 6G [1–3]. The trend for each generation is known where the 'odd' generations (1G, 3G, 5G) have been regarded as evolutions while the even generations (2G, 4G) might be perceived as revolutions for the communication world, therefore, it is the perfect time to envision a revolution with 6G.

Future 6G applications/use-cases demand solutions to combat challenging communication engineering dilemmas. Several of these applications include truly immersive virtual/augmented reality, holographic communication, remote education and much more [4]. These applications require high reliability, data rates and security all in conjunction with low latency. Future wireless networks, particularly at the physical layer, require unique approaches in terms of flexibility, novel waveforms and techniques to satisfy its stringent requirements.

In light of the above mentioned, to tackle some of these 6G demands, there has been a growing interest among the telecommunication community in manipulating and controlling the wireless propagation environment. Reconfigurable intelligent surface (RIS)-empowered communications is currently a hot topic with the aim of improving signal quality and dynamically reconfiguring the wireless channel [5–9]. RIS has emerged as a cost-effective solution to manipulate the propagation environment with its unique and effective functionalities such as controllable reflecting, absorbing and shifting phases without buffering or processing the incoming signals, hence, it is qualified as a passive component [10].

Recently, numerous noteworthy studies have been carried out demonstrating the potential of RIS technologies towards 6G wireless networks. In the literature, RISs have been integrated with existing technologies such as orthogonal frequency division multiplexing (OFDM) and multiple-input multiple-output (MIMO) to enhance their performance. Solutions have been proposed for fast and efficient channel estimation, maximizing the average sum-rate over subcarriers and maximizing the downlink achievable rate for the user by jointly optimizing the transmit power allocation at the base station (BS) and the passive reflection coefficients at the RIS [11–17]. The RIS concept has also been investigated to work collectively with related approaches, such as relaying and backscatter communications. Flexible and cost-/power-effective hybrid transmission schemes involving a relay and a passive RIS are presented in [18] while [19] investigates the limitation of phase shifts for practical RIS systems. In addition, fair comparisons have been made between relays and RISs based on signal-to-interference-noise ratio (SINR) maximization for different scenarios such as varying pathloss models and RIS positionings [20]. Furthermore, RIS channel models have been investigated thoroughly for different scenarios such as indoor and outdoor applications and their benefits have been emphasized [21, 22].

In general, RIS literature mostly tends to focus on maximizing the sum-rate/achievable rate of a wireless communication system by adjusting the reflection phases of incoming signals to the RIS, hence, improving the signal-to-noise ratio (SNR) at the target receiver or receivers [23]. However, there have been other captivating studies with different approaches such as investigating the operation of an RIS under predictable mobility considering Doppler effect and delay spread [24], feasibility of mitigating multipath fading and Doppler effects stemming from mobile receivers [25], optimizing the orientation and distance of the RIS from the users and base station to maximize cell coverage [26] and so on [27–32]. In general, studies in the literature mostly consider flat-fading systems, and frequency-selective fading is not sufficiently investigated within the context of RIS systems except for some promising OFDM-based RIS solutions. Conventionally, for systems without OFDM, equalization is applied at the receiver (Rx) to combat frequency selectivity; however, it is not always efficient. For example, numerous systems, such as Internet-of-Things (IoT) devices, will be connected wirelessly in the future and it is not efficient for them to implement complex equalization algorithms. Thus, over-the-air equalization with a passive RIS is an efficient alternative for these devices requiring equalization. To the best of the authors' knowledge, existing studies in the literature severely lack a thorough analysis of equalization methods with the aid of an RIS. Specifically, inter-symbol interference (ISI) caused by multipath fading has been reduced by optimizing the phase shifts at the RIS with the help of an iterative algorithm very recently [33]. However, this study does not provide a theoretical background or a probabilistic analysis on the elimination of ISI. Nonetheless, a unified framework for multipath fading mitigation as well as a comprehensive analysis on the potential effects of ISI are still missing in the open literature.

Against this background, this paper analyses an RIS-aided wireless network and follows a comprehensive signal processing analysis to show that conventional RIS phase alignment schemes may be leveraged to equalize the channel over-the-air. In other words, it is shown that the considered system transforms the frequency-selective channel between the transmitter (Tx) and the RIS and/or RIS and the receiver (Rx) to frequency-flat, resulting in a end-to-end frequency-flat channel from the perspective of the Rx. Specifically, the RIS adjusts its phases to eliminate only the phase terms of the strongest tap from the propagation channel. Thus, only the magnitude of the strongest tap has been boosted in a clever manner using the RIS while the other non-optimized taps are regarded as ISI terms. In other words, the effect of the ISI terms are alleviated by maximizing the desired tap of the channel response. In the literature, existing studies amplify the strongest tap to enhance communication performance, however, this study primarily aims to analyse and show that the ISI is also eliminated with the help of the RIS. We show that when the desired tap becomes at least 10 times greater in terms of squared magnitude, the ISI terms become negligible, hence, reliable transmission will be provided by suppressing the effect of ISI terms. Unlike most of the studies in the RIS literature, which aim to maximize the sum-rate, capacity or other kind of performance indicators, our primary concern is to exploit the RIS from the perspective of ISI minimization. Rather than proposing a new system, our idea is based on the existing RIS systems subject to the frequency-selective fading channels. The considered system is practically better operated in small devices in which multi-carrier waveforms cannot be utilized to combat with ISI. The contributions of this study can be summarized and highlighted as follows:

- A unified framework is proposed to virtually eliminate the frequency-selectivity of the end-to-end system by alleviating the effect of ISI resulting from multipath components.
- A thorough end-to-end system model of the proposed scheme from continuous-time to discrete-time domain is provided and analysed along with computer simulations.
- The ISI elimination probability is calculated for different configurations using a probabilistic approach.
- A theoretical error probability analysis is provided along with computer simulation results for varying parameters and scenarios.

The rest of the paper can be summarized as follows. In Section 2, the generalized end-to-end system model of the proposed scheme is presented. In Section 3, an analysis of the probability of ISI elimination for different scenarios is conducted. Section 4 provides the error probability analysis of the system and in Section 5, computer simulation results are given. Finally, in Section 6 the paper is concluded with the final remarks.

## 2 | END-TO-END SYSTEM MODEL

In this section, we consider a linear time-invariant (LTI) and wideband communication system in the presence of an RIS and present our end-to-end system model. For traceability, a user-friendly introduction in terms of signal processing steps is provided and we progress step by step to illustrate the affect of the RIS for a frequency-selective channel.

In wideband communication systems, the symbol time $T$ is shorter than the inverse of coherence bandwidth, that is, the delay spread. Since the multipath delay spread is larger than the symbol time, different multipath components interfere with each other. Therefore, in general, the multipath components of this system are resolvable and the difference of the delays between the components significantly exceeds the symbol duration. This interference results in a well-known distortion called ISI which makes the channel frequency-selective. The goal of the proposed system is to virtually eliminate ISI via an RIS and transform the end-to-end frequency-selective channel to frequency-flat, that is, by accomplishing some sort of virtual equalization over-the-air.

Under the LTI assumption, we can express a general communication system's received continuous-time signal $y(t)$ in the absence of noise using the convolution integral as

$$y(t) = \int_{-\infty}^{\infty} h_c(\tau) x(t-\tau) d\tau, \quad (1)$$

where $h_c(t)$ is the end-to-end channel impulse response, $x(t)$ is the transmitted signal and $\tau$ is the delay term. It is possible to transition to discrete-time by sampling the continuous-time signal $y(t)$, and rewriting the communication signal model as

$$y[n] = \sum_{k=-\infty}^{\infty} h[k] x[n-k]. \quad (2)$$

In baseband, the transmitted signal $x(t)$, with bandwidth $B/2$ Hz, is band-limited, that is, the input to an LTI-type communication system with impulse response $h_c(t)$. Since the input is band-limited, only the portion of $h_c(t)$ that exists in the bandwidth of $x(t)$ is important to us. We can obtain the continuous-time complex baseband channel as [34, eq. (3.348)]

$$h(t) = B \int \text{sinc}(B(t-\tau)) h_c(\tau) e^{j2\pi f_c \tau} d\tau, \quad (3)$$

where $B$ and $f_c$ stand for the complete bandwidth of the ideal lowpass filter and the operating frequency, respectively. In this section, we will consider three scenarios of frequency-selectivity: Tx-RIS is frequency-selective while RIS-Rx is frequency-flat, Tx-RIS is frequency-flat while RIS-Rx is frequency-selective and when the Tx-RIS and RIS-Rx channels are both frequency-selective. As illustrated in Figure 1, an outdoor and RIS-assisted communication system is considered under the blocked line-of-sight (LOS) path between the Tx and Rx. Here, we consider the most general case where the channel from the Tx to the RIS is frequency-selective and from the RIS to the Rx is also frequency-selective. When the RIS is near the Tx or Rx, which is a common assumption in the literature, the corresponding channel may be considered frequency-flat due to the close proximity of the RIS to the terminals. It should be noted that the analysis when the RIS is close to the Rx also applies to the case when the RIS is close to the Tx. Finally, we provide an additional discussion by considering the direct Tx–Rx link in Section 3.4. The RIS is equipped with $M$ passive reflecting elements that may be adjusted according to the wireless channel. In Figure 1, a general system with $N_h$ paths from the Tx to the RIS and $N_g$ paths from the RIS to the Rx is presented. We assume a frequency-selective channel between both the Tx-RIS and RIS-Rx, that is, all the multipath components $N_h$ and $N_g$ are resolvable. The continuous-time impulse response of the end-to-end channel in Figure 1 can be given as

$$h_c(t) = \sum_{l_1=1}^{N_h} \sum_{l_2=1}^{N_g} \sum_{m=1}^{M} \alpha_{l_1} \beta_{l_2} \delta(t - \tau_{l_1}^m - \xi_{l_2}^m) e^{j\theta_m}, \quad (4)$$

where $\alpha_{l_1}$, $\beta_{l_2}$, $\tau_{l_1}^m$, $\xi_{l_2}^m$ represent the complex channel coefficients and delays of the $l_1$th and $l_2$th path through the $m$th RIS element, corresponding to the Tx-RIS and RIS-Rx channels, respectively. Furthermore, $e^{j\theta_m}$ represents the adjustable phase term for the $m$th RIS element. Since the RIS acts as an LTI filter, $e^{j\theta_m}$ is not dependent on time for one symbol duration.

The continuous-time channel response in (3) can be written in discrete form as follows [34, eq. (3.386)]:

$$h[k] = Th(kT), \quad (5)$$

where $T = \frac{1}{B}$. By applying an ideal lowpass filter, with sinc-type Nyquist pulse shaping, and the same bandwidth as the input signal similarly from (3), we may simplify (4) to discrete form by using (5) and using [34, eq. (3.385)] obtain

$$h[n] = \int \text{sinc}(n - B\tau) h_c(\tau) e^{-j2\pi f_c \tau} d\tau =$$

$$\sum_{l_1=1}^{N_h} \sum_{l_2=1}^{N_g} \sum_{m=1}^{M} \alpha_{l_1} \beta_{l_2} e^{j\theta_m} \times \text{sinc}(n - B(\tau_{l_1} + \xi_{l_2})) e^{-j2\pi f_c(\tau_{l_1}^m + \xi_{l_2}^m)}. \quad (6)$$

It should be noted that since the delay differences coming from the RIS elements are insignificant compared to symbol duration, the $(\cdot)^m$ terms stemming from (4) inside the sinc function in (6) are removed and the delay for the first RIS element is used for all RIS elements. In (6), $\tau_{l_1} + \xi_{l_2}$ may be a fraction of a symbol period resulting in an impairment where the receiver does not sample precisely at the right time. Therefore, to counter this impairment, symbol synchronization is employed at the Rx to fix the sample timing error by enabling the delays $\tau_{l_1} + \xi_{l_2}$ to become integer fractions of the symbol duration [34]. When symbol synchronization is applied at the receiver, (6) can be re-expressed as follows [34]:

$$h[n] = \sum_{l_1=1}^{N_h} \sum_{l_2=1}^{N_g} \sum_{m=1}^{M} h_{l_1}^m g_{l_2}^m e^{j\theta_m} \delta(n - B\tilde{\tau}_{l_1} - B\tilde{\xi}_{l_2}), \quad (7)$$

where $\tilde{\tau}$ and $\tilde{\xi}$ represent the non-fractional (integer) delays of the paths after symbol synchronization. It is worth noting that the $\delta(\cdot)$ in (7) is the Kronecker delta function while in (4) is direc delta. Additionally, the delay terms in the exponential of (6) are now combined with the channel coefficients for the $m$th

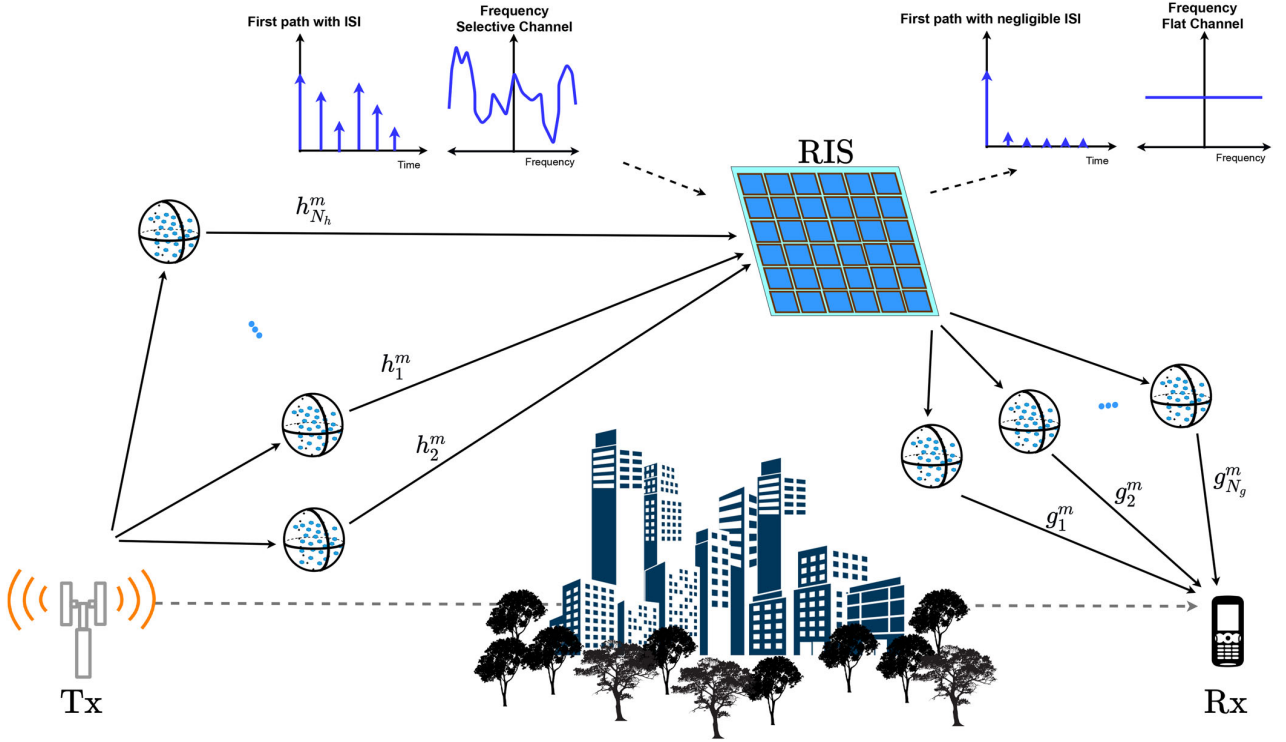

**FIGURE 1** Proposed over-the-air equalization system model with $N_h$-paths between Tx-RIS and $N_g$ paths between the RIS-Rx

element of each path, which are denoted as

$$h_{l_1}^m = \alpha_{l_1} e^{-j2\pi f_c \tau_{l_1}^m},$$
$$g_{l_2}^m = \beta_{l_2} e^{-j2\pi f_c \xi_{l_2}^m}. \quad (8)$$

Hereby, the complex baseband received signal $y(t)$ may be written in discrete form as in (2), by sampling at integer multiples of $T$ as $y[n] = y(nT)$. The received signal corrupted with additive white Gaussian noise (AWGN) samples $w[n]$, where $w \sim \mathcal{CN}(0, W_0)$ with $\mathcal{CN}(0, \sigma^2)$ denoting complex Gaussian distribution with variance $\sigma^2$, may be expressed as

$$y[n] = \sum_{l_1=1}^{N_h} \sum_{l_2=1}^{N_g} \sum_{m=1}^{M} h_{l_1}^m g_{l_2}^m e^{j\theta_m} x[n - B\tilde{\tau}_{l_1} - B\tilde{\xi}_{l_2}] + w[n]. \quad (9)$$

For simplicity, throughout the paper we denote the strongest tap with the first tap where $(l_1, l_2) = (1, 1)$. If we arrange the first path term and other terms separately, which will be treated as ISI later on, the received signal can be reorganized as

$$y[n] = \sum_{m=1}^{M} h_1^m g_1^m e^{j\theta_m} x[n - B\tilde{\tau}_1 - B\tilde{\xi}_1]$$
$$+ \sum_{\substack{l_1=1 \\ (l_1, l_2) \neq (1,1)}}^{N_h} \sum_{l_2=1}^{N_g} \sum_{m=1}^{M} h_{l_1}^m g_{l_2}^m e^{j\theta_m} x[n - B\tilde{\tau}_{l_1} - B\tilde{\xi}_{l_2}] + w[n]. \quad (10)$$

Even when the symbol timing has been corrected, another impairment occurs with larger delays due to an unknown propagation delay, which is a multiple of the symbol period. These integer offsets create a mismatch between the indices of the transmitted and received symbols, and the correction of this impairment requires frame synchronization [34]. After frame synchronization is applied, the discrete-time received signal in (10) is rewritten as

$$y[n] = \sum_{m=1}^{M} h_1^m g_1^m e^{j\theta_m} x[n] + \sum_{\substack{l_1=1 \\ (l_1, l_2) \neq (1,1)}}^{N_h} \sum_{l_2=1}^{N_g} \sum_{m=1}^{M} h_{l_1}^m g_{l_2}^m e^{j\theta_m}$$
$$\times x[n - ((B\tilde{\tau}_{l_1} + B\tilde{\xi}_{l_2}) - (B\tilde{\tau}_1 + B\tilde{\xi}_1))] + w[n], \quad (11)$$

where $h_1^m g_1^m = |h_1^m||g_1^m|e^{j\phi_m}$, $h_{l_1}^m g_{l_2}^m = |h_{l_1}^m||g_{l_2}^m|e^{j\psi_m^{l_1,l_2}}$ and $|\cdot|$ is the absolute value operator. Hence, we may express the SINR $\gamma$ of the system as follows:

$$\gamma = \frac{E_s \left| \sum_{m=1}^{M} |h_1^m||g_1^m|e^{j\phi_m} e^{j\theta_m} \right|^2}{E_s \sum_{\substack{l_1=1 \\ (l_1, l_2) \neq (1,1)}}^{N_h} \sum_{l_2=1}^{N_g} \left| \sum_{m=1}^{M} |h_{l_1}^m||g_{l_2}^m|e^{j\psi_m^{l_1,l_2}} e^{j\theta_m} \right|^2 + W_0}. \quad (12)$$

Here $E_s$ is the energy of the transmitted signal, $\phi_m = \angle(h_1^m g_1^m)$, and $\psi_m^{l_1,l_2} = \angle(h_{l_1}^m g_{l_2}^m)$, where $\angle(\cdot)$ denotes the phase of a complex term. By adjusting the RIS phases according to the first tap as $\theta_m = -\phi_m$, the phase terms of the numerator are eliminated and $\gamma$ is simplified to

$$\gamma = \frac{E_s(\sum_{m=1}^{M} |h_1^m||g_1^m|)^2}{E_s \sum_{\substack{l_1=1 \\ (l_1,l_2)\neq(1,1)}}^{N_h} \sum_{l_2=1}^{N_g} \left|\sum_{m=1}^{M} |h_{l_1}^m||g_{l_2}^m| e^{j\psi_m^{l_1,l_2}} e^{j\theta_m}\right|^2 + W_0}. \quad (13)$$

Here we aim to arrange $e^{j\theta_m}$ specifically to cancel out the phase terms of the first path $\phi_m$, that is, the phase of the numerator. Hence, the numerator of the SINR becomes real valued while the denominator with the ISI components are complex and deteriorated with the additional random phases. In this way, we can maximize the first and strongest tap of the system making it several times more powerful in terms of magnitude than the combined ISI terms, which will be further analysed in Section 3 to eliminate the ISI effect.

## 3 | ISI ELIMINATION: A PROBABILISTIC APPROACH

In this section, the ISI elimination concept introduced in Section 2 is analysed with a detailed probabilistic approach via computer simulation results. Two scenarios of an RIS system are analysed. The first is when the RIS is close to the Tx or Rx resulting in the Tx-RIS or RIS-Rx link to be frequency-flat while the other is frequency-selective, characterized by Scenario 1. On the other hand, the latter case is when both Tx-RIS and RIS-Rx links of the RIS are frequency-selective (Scenario 2).

### 3.1 | Scenario 1

In this subsection, we assume only one path from the RIS to the Rx ($N_g = 1$) since the corresponding channel experiences flat fading due to close positioning of the RIS to the Rx. It should be noted that for the case of ($N_h = 1$), the same results in this section are valid. We remove the underscore and denote the single path from the RIS-Rx simply as $g^m$. The channel fading coefficients of the Tx-RIS path, $h_{l_1}^m$ where $l_1 = 1, \ldots, N_h$ and the single path from the RIS to the Rx, $g^m$, reflecting from the $m^{th}$ RIS element $m = 1, \ldots, M$ are independent and identically distributed under the Rayleigh fading assumption for a worst-case scenario analysis. Hence, these channel coefficients follow the distribution $h_{l_1}^m, g^m \sim \mathcal{CN}(0, \sigma_i^2)$, where $\sigma_i^2$ and $i \in \{h, g\}$ is a function of the large-scale path loss. We may express the end-to-end channels for the first path and the ISI paths, respectively, as $A$ and $B$:

$$A = \sum_{m=1}^{M} h_1^m e^{j\theta_m} g^m = \sum_{m=1}^{M} |h_1^m||g^m|,$$

$$B = \sum_{l_1=2}^{N_h} b_{l_1} = \sum_{l_1=2}^{N_h} \sum_{m=1}^{M} h_{l_1}^m e^{j\theta_m} g^m, \quad (14)$$

where $e^{j\theta_m} = e^{-j(\angle h_1^m + \angle g^m)}$. Since $h_1^m$ and $g^m$ are independently Rayleigh distributed random variables, the mean and the variance of their product are $E[|h_1^m||g^m|] = \frac{\sigma_{h_1}\sigma_g \pi}{4}$ and $\text{VAR}[|h_1^m||g^m|] = \sigma_{h_1}^2 \sigma_g^2 (1 - \frac{\pi^2}{16})$, respectively. When the number of RIS elements $M$ is significantly large, from the Central Limit Theorem, $A$ converges to a real Gaussian distribution random variable with $A \sim \mathcal{N}(\frac{M\pi\sigma_{h_1}\sigma_g}{4}, M\sigma_{h_1}^2 \sigma_g^2 (1 - \frac{\pi^2}{16}))$. On the other hand, since the phase terms exist in the ISI paths, $b_{l_1}$ converges to a complex Gaussian random variable. The mean and variance of the product of the terms in $b_{l_1}$ are calculated as $E[h_{l_1}^m e^{j\theta_m} g^m] = 0$ and $\text{VAR}[h_{l_1}^m e^{j\theta_m} g^m] = M\sigma_{h_{l_1}}^2 \sigma_g^2$ resulting in $B \sim \mathcal{CN}(0, M(N_h - 1)\sigma_{h_{l_1}}^2 \sigma_g^2)$.

As it can be seen from (14), $A$ and the $B$ have $g^m$ as a common term; however, they are independent since they are uncorrelated and Gaussian distributed. To find their correlation coefficient, the covariance of $A$ and $B$ is required and since $B$ has zero mean, it may be expressed as

$$E[AB] = E\left[\left(\sum_{m=1}^{M} h_1^m e^{j\theta_m} g^m\right)\left(\sum_{l_1=2}^{N_h} \sum_{m=1}^{M} h_{l_1}^m e^{j\theta_m} g^m\right)\right]. \quad (15)$$

After adjusting the RIS phases according to the first path, (15) may be re-expressed as

$$E[AB] = E\left[\left(\sum_{m=1}^{M} |h_1^m||g^m|\right)\left(\sum_{l_1=2}^{N_h} \sum_{m=1}^{M} h_{l_1}^m e^{-j\angle h_1^m} |g^m|\right)\right]. \quad (16)$$

Here, $b_{l_1}$ is a complex Gaussian term with zero mean hence, $E[AB] = 0$. Therefore, the correlation coefficient of $A$ and $B$ is also zero resulting in $A$ and $B$ being uncorrelated. Furthermore, they are Gaussian distributed and independent.

We assume that when $A$ is at least 10 times greater in squared magnitude than the power of the ISI terms, the ISI terms will be negligible and can be ignored. To generalize this, the probability that $A$ is $X$ times greater in squared magnitude than the sum of $b_{l_1}$ terms, where $b_{l_1} = \sum_{m=1}^{M} h_{l_1}^m e^{-j\angle h_1^m} |g^m|$ also converges to a complex Gaussian random variable, may be expressed as follows:

$$P\left(A^2 - X \sum_{l_1=2}^{N_h} |b_{l_1}|^2 > 0\right) = 1 - P(C < 0) =$$
$$1 - \int_{-\infty}^{0} f_C(c) dc, \quad (17)$$

where $C = A^2 - X \sum_{l_1=2}^{N_h} |b_{l_1}|^2$, and $f_C(c)$ is the probability density function (PDF) of $C$. Therefore, to ensure successful ISI elimination, we assume that $X$ would be at least 10. In (17), $A^2$ and $\sum_{l_1=2}^{N_h} |b_{l_1}|^2$ follow non-central chi-square distribution with one degree of freedom and central chi-square distribution with $2(N_b - 1)$ degrees of freedom, respectively. Here, $C$ is the difference of independent non-central and central chi-squared random variables [35]. Thus, we would need to integrate the PDF of the difference of non-central chi-squared and central chi-squared random variables accordingly. However, the PDF is expressed in the form of infinite series in Whittaker functions where themselves are expressed in terms of the confluent hypergeometric function. Due to the complexity of the PDF calculation and the absence of functions in standard mathematical software as mentioned in [35], it is extremely complicated to approach this method. On the other hand, the characteristic function (CF) of $C$ is simply derived and expressed as:

$$\Psi_C(\omega) = \frac{1}{(1 - 2j\omega\sigma_1^2)^{0.5}(1 + 2j\omega\sigma_2^2)^{(N_b-1)}} \exp\left(\frac{j\omega\mu_1^2}{1 - 2j\omega\sigma_1^2}\right), \quad (18)$$

where $\mu_1^2, \sigma_1^2$ and $\sigma_2^2$ stand for the non-centrality parameter which is the mean squared of $A$, variance of $A$ and variance of $\sqrt{X}B$, respectively. Since $M$ is directly multiplied with a power of $\sigma_i$, in the statistics provided for $A$ and $B$, for a small $\sigma_i$ the number of RIS elements $M$ is not expected to significantly affect the result. Using the Gil-Pelaez's inversion formula [36], it is possible to obtain the cumulative distribution function (CDF) of this difference of chi-squared random variables as

$$F_C(c) = \frac{1}{2} - \int_0^\infty \frac{\Im\{e^{-j\omega c}\Psi_C(\omega)\}}{\omega\pi} d\omega, \quad (19)$$

where $\Im$ denotes the imaginary part, $F_C(c) = P(C \leq c)$ is the CDF and $\Psi_C(\omega)$ is the CF of $C$. Solving for $c = 0$ and using numerical integration, the probability of $A$ being $X$ times greater than $B$ may be obtained.

### 3.2 | Scenario 2

In this subsection, the scenario where the channels from the Tx-RIS and RIS-Rx are both frequency-selective is taken into consideration. Similar to the previous case, the end-to-end channels for the first path and the ISI paths may be represented as

$$A = \sum_{m=1}^{M} h_1^m e^{j\theta_m} g_1^m = \sum_{m=1}^{M} |h_1^m||g_1^m|,$$

$$B = \sum_{l_1=2}^{N_b} \sum_{l_2=2}^{N_g} b_{l_1,l_2} = \sum_{l_1=2}^{N_b} \sum_{l_2=2}^{N_g} \sum_{m=1}^{M} h_{l_1}^m e^{j\theta_m} g_{l_2}^m, \quad (20)$$

where $e^{j\theta_m} = e^{-j(\angle h_1^m + \angle g_1^m)}$. The mean and the variance of $A$ remains the same and $A$ has the same distribution. On the other hand, the mean and variance of $b_{l_1,l_2}$ can be recalculated as $E[h_{l_1}^m e^{j\theta_m} g_{l_2}^m] = 0$ and $VAR[h_{l_1}^m e^{j\theta_m} g_{l_2}^m] = M\sigma_{h_{l_1}}^2 \sigma_{g_{l_2}}^2$ resulting in $B \sim \mathcal{CN}(0, M(N_g - 1)(N_b - 1)\sigma_{h_{l_1}}^2 \sigma_{g_{l_2}}^2)$. $A$ and $B$ remain uncorrelated which can be seen by reapplying (15) and (16) to obtain

$$E[AB] = E\left[\left(\sum_{m=1}^{M} |h_1^m||g_1^m|\right)\left(\sum_{l_1=2}^{N_b} \sum_{l_2=2}^{N_g} \sum_{m=1}^{M} h_{l_1}^m e^{-j\angle\theta_m} g_{l_2}^m\right)\right], \quad (21)$$

the probability that $A$ is $X$ times greater in squared magnitude than $B$ may be expressed as follows:

$$P\left(A^2 - X \sum_{l_1=2}^{N_b} \sum_{l_2=2}^{N_g} |b_{l_1,l_2}|^2 > 0\right) = 1 - P(C < 0)$$

$$= 1 - \int_{-\infty}^{0} f_C(c) dc, \quad (22)$$

where $C = A^2 - X \sum_{l_1=2}^{N_b} \sum_{l_2=2}^{N_g} |b_{l_1,l_2}|^2$. In (22), $A^2$ and $\sum_{l_1=2}^{N_b} \sum_{l_2=2}^{N_g} |b_{l_1,l_2}|^2$ follow non-central chi-square distribution with one degree of freedom and central chi-square distribution with $2(N_b - 1)(N_g - 1)$ degrees of freedom, respectively. Due to the complexity of the resulting PDF, we proceed with the Gil-Pelaez's inversion formula and the CF once more. The CF of $C$ may be expressed as follows:

$$\Psi_C(\omega) = \frac{1}{(1 - 2j\omega\sigma_1^2)^{0.5}(1 + 2j\omega\sigma_2^2)^{(N_b-1)(N_g-1)}} \exp\left(\frac{j\omega\mu_1^2}{1 - 2j\omega\sigma_1^2}\right), \quad (23)$$

where $\mu_1^2, \sigma_1^2$ and $\sigma_2^2$ stand for the non-centrality parameter which is the mean square of $A$, variance of $A$ and variance of $\sqrt{X}B$, respectively. Using (19), we may obtain the CDF $F_C(c) = P(C \leq c)$, solve for $c = 0$ and calculate the probability of $A$ being at least 10 times greater than $B$ using numerical integration.

### 3.3 | ISI elimination for Scenarios 1 and 2

Computer simulation results in this section provide a clear visualization of the probability of the strongest path being $X$ times greater in squared magnitude than the combined ISI paths allowing us to clearly express that they are negligible and eliminated. Both scenarios of ISI elimination are discussed.

In Figure 2a, we provide the ISI elimination probability for Scenario 1 for varying $N_b$ at $M = 256$ and $\sigma_i = 10^{-2}$. It should be noted that for practical values of $\sigma_i < 10^{-3}$ we observe very high probabilities of ISI elimination and purposefully increased $\sigma_i$ for observational purposes. It can be seen that under the worst-case scenario with a uniform power-delay-profile (PDP),

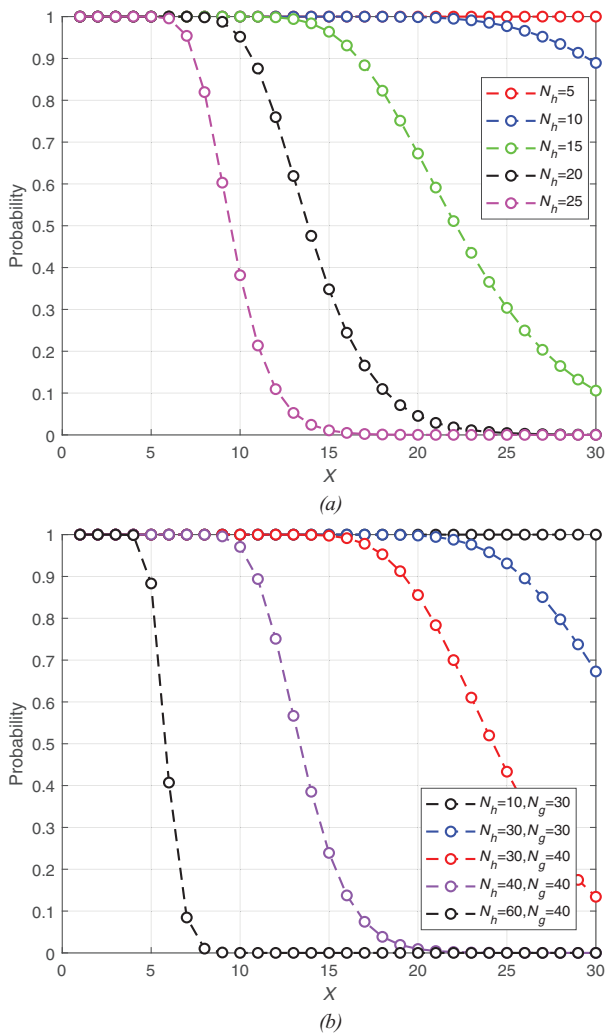

**FIGURE 2** Probability of RIS-induced first tap channel coefficient being $X$ times greater in squared magnitude than ISI term for $M = 256$ and varying $N_h$ for (a) Scenario 1 where $\sigma_i = 10^{-3}$ and (b) Scenario 2 where $\sigma_i = 10^{-2}$

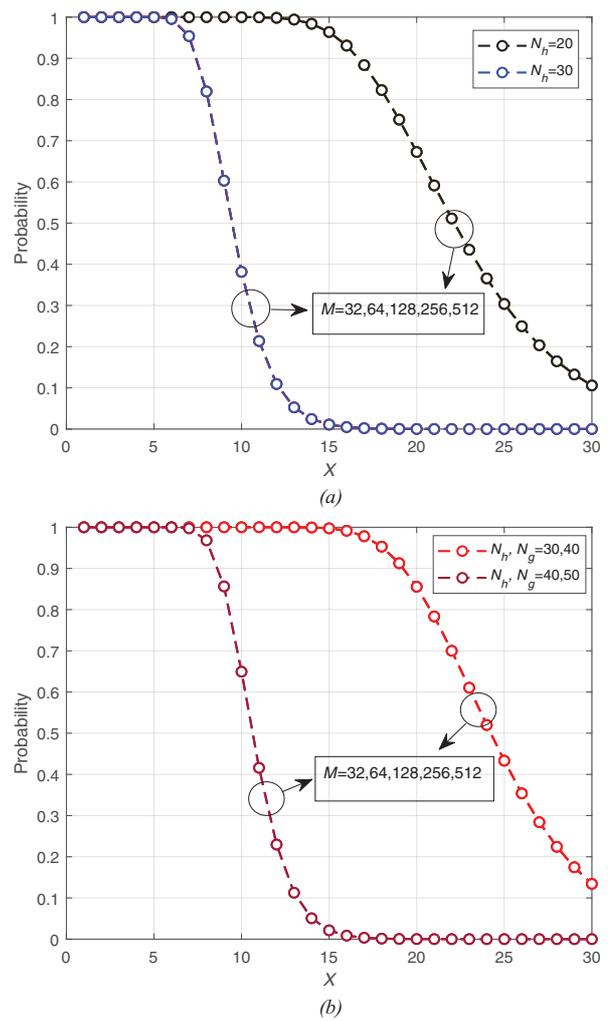

**FIGURE 3** Probability of RIS-induced first tap being $X$ times greater in squared magnitude than ISI term for varying $M$ under (a) Scenario 1 where $\sigma_i = 10^{-3}$ and (b) Scenario 2 where $\sigma_i = 10^{-2}$

high $\sigma_i$ and Rayleigh distribution, the probability of the first path being at least $X = 10$ times greater in squared magnitude than the ISI paths is about 99% at $N_h = 15$. In general, Figure 2a demonstrates that for varying total number of paths ($N_h$), that is, $N_h - 1$ ISI paths, the probability that the first tap is 10 times greater in squared magnitude than the combined ISI terms is very high for even severe conditions. It can also be seen that not only is it 10 times greater, the first tap is more than 30 times greater in squared magnitude than the combined ISI paths for many of cases. In Figure 2b, we vary both $N_h$ and $N_g$ for $M = 32$ and $\sigma_i = 10^{-3}$. Even with frequency selectivity at both links of the RIS and numerous ISI paths, for practical values of $\sigma_i$ the ISI is completely eliminated. For a severe number of ISI taps at high fading power, we can still see that the ISI is eliminated 98% of the time for $N_h = 40$, $N_g = 40$. Even though the ISI is almost completely negligible in scenarios with smaller $\sigma_i$, we can see that as the number of ISI terms increases, the probability of the ISI elimination degrades. However, the probability of ISI elimination for even a large number of ISI paths is high.

In Figure 3, $N_h$ and $N_g$ are set to constant values for both scenarios of frequency selectivity to examine the performance of varying $M$ values, that is, RIS sizes. It can be seen that for a small $\sigma_i$, the probability of ISI elimination is unchanged with the RIS size, because as $M$ increases, the results stay the same and the curves fully overlap each other for all scenarios along with cases of $N_h$ and $N_g$. This goes for any RIS size and number of paths. As mentioned previously, this is because in the statistics of $A$ and $B$, $M$ is multiplied with $\sigma_i$ values thus, for a small $\sigma_i$, the effect of $M$ in the probability of elimination is greatly reduced. On the other hand, if $\sigma_i$ were to be a very large value which is not practical, a change in $M$ may impact the results and the curves in Figure 3 would not overlap.

It should be noted that for all computer simulations, we assume a uniform PDP, unless stated otherwise, which models a worst-case scenario, hence, for other PDPs the results would be superior. Additionally, for a smaller $\sigma_i$, the probability of negligible ISI approaches even closer to unity. We can conclude that even in the worst-case scenario, we can eliminate the ISI and maximize the squared magnitude of the first tap. For practical values of $\sigma_i$, we were not able to see an effect on changing the ISI paths or RIS elements because we would obtain perfect ISI elimination thanks to the RIS operation, hence for observational purposes we provided worst-case scenarios. These probabilities remain outstanding and ensure that the ISI is almost negligible, which will be further verified in terms of error performance analysis in the next section.

### 3.4 | General framework for ISI elimination

In this subsection, we assume there is a direct path from the Tx–Rx, hence, ISI elimination would need to be approached with caution and alternative perspectives. When there is a direct path from the Tx–Rx, the received discrete-time baseband signal may be expressed as:

$$y[n] = \sum_{d=1}^{D} h_d x[n - B\tilde{\xi}_d] + \sum_{l_1=1}^{N_b} \sum_{l_2=1}^{N_g} \sum_{m=1}^{M} h_{l_1}^m g_{l_2}^m e^{j\theta_m}$$
$$\times x[n - (B\tilde{\tau}_{l_1} + B\tilde{\xi}_{l_2})] + w[n], \quad (24)$$

where $D$ is the number of direct paths from the Tx–Rx, $h_d$ is the channel coefficient for the $d$th direct path and $\tilde{\xi}$ is the corresponding integer delay. In this case, there exists multiple PDPs. The received signals from the RIS may arrive at a different time than the direct signal paths to the Rx which can be an ameliorating or deteriorating factor depending on the scenario. Assuming the strongest and fastest tap, where the phases are arranged by the RIS coherently, are aligned and non-resolvable with the first tap, thus, ending up in the same delay bin. Therefore, the RIS may be exploited to reduce ISI that may be affecting the direct path by boosting the strongest signal of the direct path. However, this time alignment is not a guarantee in practical scenarios and the RIS may instead cause additional ISI to the direct signal if the PDP taps are resolvable with the direct path taps. In such a case, the RIS may try to minimize the ISI taps through a sort of complex optimization problem. Another alternative may be to activate a full absorption mode on the RIS and possibly reduce ISI by eliminating signals that may cause interference to the dominant direct signal. These will be our future research directions to combat multipath fading comprehensively.

## 4 | THEORETICAL ERROR PROBABILITY ANALYSIS

In this section, the system of Figure 1 is revisited, where the transmission is carried out via an RIS under a blocked link between the Tx and Rx. For the sake of simplicity and to provide a clear insight, throughout the rest of this paper we consider a single side frequency-selective scenario (Scenario 1). For this setup, $d_{\text{Tx-Rx}}$, $d_{\text{Tx-RIS}}$, $d_{\text{RIS-Rx}}$, represent the distances between the Tx and Rx, Tx and RIS, RIS and Rx, respectively. In addition, $h_{\text{Tx}}$, $h_{\text{RIS}}$, $h_{\text{Rx}}$, and $f_c$, represent the height of the Tx, RIS, Rx and the operating frequency, respectively. The 3GPP UMi point-to-point NLOS path loss model for a single Tx/Rx path is also considered [37]. Within the $2 - 6$ GHz frequency band and Tx–Rx distance ranging from $10 - 2000$ m, the 3GPP UMi path loss model for NLOS transmission is expressed as:

$$L(d)[\text{dB}] = 36.7\log_{10}(d) + 22.7 + 26\log_{10}(f_c). \quad (25)$$

The fading channel between the single antenna Tx and the $m$th RIS element, and $m$th RIS element to the single antenna Rx are denoted as $h_{l_1}^m$ and $g^m$, respectively. It is assumed that Rayleigh fading channels $h_{l_1}^m$ and $g^m$ follow $\mathcal{CN}(0, \sigma_i^2)$, distribution where $\sigma_i^2$ is the variance of the complex Gaussian distribution. The RIS is composed of $M$ passive and controllable reflecting elements. Revisiting Section 2 with $N_g = 1$, the received signal can be written as

$$y[n] = \sum_{m=1}^{M} \alpha_1^m \beta_1^m e^{j\theta_m} x[n] + \sum_{l_1=2}^{N_b} \sum_{m=1}^{M} \alpha_{l_1}^m e^{j\theta_m} \quad (26)$$
$$\times x[n - (B\tilde{\tau}_{l_1} - B\tilde{\tau}_1)] + w[n],$$

where $e^{j\theta_m}$ is the adjustable phase shift introduced by the $m$th RIS element, $x[n]$ stands for the data symbol selected from $Q$-ary phase shift keying/quadrature amplitude modulation (PSK/QAM) constellations, and $w \sim \mathcal{CN}(0, W_0)$ is the AWGN term.

With the assumption of knowledge of the channel phases at the RIS [23], it can be seen that by adjusting the $\theta_m$ term at the RIS, that is, $\theta_m = -\phi_m$ for $m = 1, \ldots, M$, the SINR can be maximized since the channel phases of the first tap is eliminated. Additionally, we assume that the ISI becomes negligible when the first tap is 10 times greater in squared magnitude than the ISI paths. Substituting (14) into (13) the maximized SINR ($\gamma_m$) can be approximated as:

$$\gamma_m = \frac{E_s A^2}{E_s \sum_{l_1=2}^{N_b} |b_{l_1}|^2 + W_0}. \quad (27)$$

Since $A^2$ is a non-central chi-squared random variable with one degree of freedom and $B = \sum_{l_1=2}^{N_b} |b_{l_1}|^2$ is central chi-squared with $2(N_b - 1)$ degrees of freedom, their SINR ratio $\gamma_m$ cannot be exactly derived and to the best of the authors' knowledge, its closed form has not been reported in the open literature. Thus, a semi-analytical approach is conducted where the distribution of $\gamma_m$ is generated via MATLAB using dfittool by carefully considering $A$ and $B$ terms.

As we may recall, $A^2$ is a chi-squared distributed random variable which is a specific case of the gamma distribution. Since the effect of the ISI term $B$ in $\gamma_m$ is negligible, we expect $\gamma_m$ to fit the gamma distribution. Figure 4a shows the PDF of $\gamma_m$

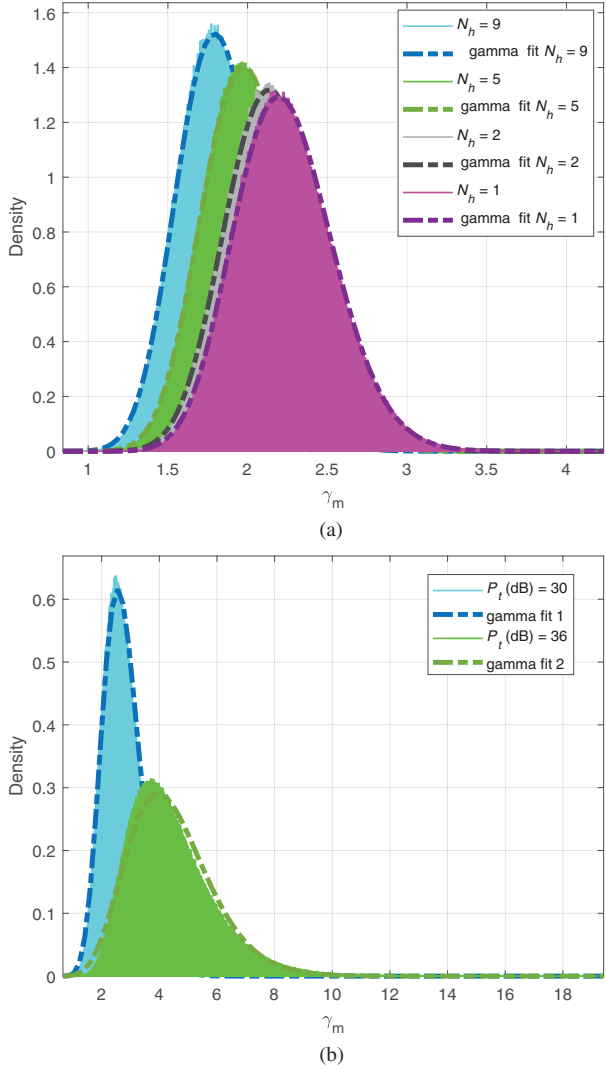

**FIGURE 4** (a) Density distribution of $\gamma_m$ for varying $N_h$ and (b) gamma distribution fit for varying $P_t$

and its fit to the gamma distribution for varying number of ISI terms ($N_h - 1$) with a fixed RIS size of $M = 128$ and $P_t = 20$ dB. Using the dfittool in MATLAB, it can be confirmed that the distribution of $\gamma_m$ fits the gamma distribution. As the number of ISI terms are increased, the PDFs of the distributions begin to differ from one another and for a large number of $N_h$, as seen from the case of $N_h = 9$, the ISI terms start to show an effect and begin to slightly deviate from an exact fit to the gamma distribution. For different parameters such as $M$, $N_h$, $P_t$, $W_0$, the gamma distribution parameters vary and a general expression cannot be derived to express the gamma distribution parameters for given specifications. On the other hand, Figure 4b shows the PDF for $M = 64$ for $P_t = 30$ dB and $P_t = 36$ dB. It can be seen that for $P_t = 30$ dB, the gamma distribution fits perfectly; however for a higher SINR where $P_t = 36$ dB, the gamma distribution does not fit perfectly. This is because as the SINR increases, the effect of the noise diminishes and the ISI terms in $B$ starts to become more effective. Thus, the distribution is affected and deviated from the chi-squared distribution, which is a form of the gamma distribution.

To be used in theoretical derivations, for each configuration, the shape $\kappa$ and scale $\rho$ parameters for the gamma distribution need to be generated. Table 1 provides the gamma distribution parameters for $M = 128$ under $W_0 = -130$ dBm noise power required to calculate the moment generating function (MGF) for different scenarios.

Specifically, the gamma distribution has the following MGF

$$M_{\gamma_m}(s) \approx (1 - \rho s)^{-\kappa}, \quad \text{for} \quad s < \frac{1}{\rho}. \qquad (28)$$

From (28), we can obtain the average SEP for $Q$-PSK signaling as in [21],

$$P_e \approx \frac{1}{\pi} \int_0^{(Q-1)\pi/Q} M_\gamma \left( \frac{-\sin^2(\pi/Q)}{\sin^2\eta} \right) d\eta, \qquad (29)$$

which for binary PSK (BPSK) simplifies to

$$P_e \approx \frac{1}{\pi} \int_0^{\pi/2} \left( 1 + \frac{\rho}{\sin^2\eta} \right)^{-\kappa} d\eta. \qquad (30)$$

## 5 | SIMULATION RESULTS

In this section, computer simulation results for the proposed over-the-air equalization system (Scenario 1) with an RIS presented. First, the bit-error rate (BER) of the system is evaluated and compared to theoretical results for varying $M$ and discussed in detail. Subsequently, the BER performance for discrete phases and ideal phases are assessed. Finally, a comparison of the proposed intelligent reflection and plain reflection, where the ISI terms remain, is presented for increasing RIS element size $M$.

For all computer simulations, path loss is included and the simulation parameters in Table 2 are used. For the BER analyses, the noise power is fixed to $-130$ dBm and the transmit power $P_t$ is varied.

In Figures 5 and 6, we present the BER results to verify the effectiveness of ISI elimination for $N_h = 2$ and varying $M$. It is clear that as the number of RIS elements $M$ increases, the performance significantly increases and less $P_t$ is needed to achieve the same error performance. However, there is a trade-off between cost and performance as $M$ increases. Specifically, from Figure 5, it can be confirmed that the ISI is negligible. At high BER values and $M$, computer simulation results exactly match the theoretical curves, which are the exact curves obtained from (30), and confirm the elimination of ISI. As seen from the case of $M = 64$ at high $P_t$ values, where the noise effect is negligible, the exact BER curves slightly differ than the simulated BER curves and this is because the negligible ISI terms beginning to present themselves which affects the distribution of $\gamma_m$ as discussed in the previous section. However, for a greater $M$, even at low BER values, the ISI is nowhere to be seen because as $M$ increases, the effect of the first tap becomes more significant. In other words, the ISI elimination probability is the same for all $M$ as we recall from Figure 3, however, as $M$ increases in the BER results, the effect of the numerator in (27) becomes more dominant than the remaining ISI terms in the denominator thus, diminishing its effect even more.

**TABLE 1** Fitted gamma distribution parameters for varying $P_t$ and ISI terms

| # of Taps | $P_t$ (dB): | 5 | 10 | 15 | 20 | 25 | 30 |
|---|---|---|---|---|---|---|---|
| $N_b = 1$ | $\kappa$ | 51.6177 | 51.6967 | 51.7189 | 51.741 | 51.7497 | 51.8369 |
| $N_b = 1$ | $\rho$ | 0.00137096 | 0.00432819 | 0.0136773 | 0.0432479 | 0.136713 | 0.431585 |
| $N_b = 2$ | $\kappa$ | 51.6532 | 51.7883 | 51.8742 | 50.9644 | 41.8229 | 19.6424 |
| $N_b = 2$ | $\rho$ | 0.00136879 | 0.00430832 | 0.013516 | 0.0427256 | 0.156152 | 0.921367 |
| $N_b = 5$ | $\kappa$ | 51.7566 | 52.0687 | 52.3222 | 49.2235 | 31.7999 | 13.2366 |
| $N_b = 5$ | $\rho$ | 0.00136239 | 0.0042491 | 0.013054 | 0.0409241 | 0.166368 | 0.845089 |
| $N_b = 9$ | $\kappa$ | 51.8945 | 52.4418 | 52.9093 | 48.0063 | 28.909 | 14.2753 |
| $N_b = 9$ | $\rho$ | 0.00135392 | 0.0041721 | 0.0124799 | 0.0381424 | 0.145555 | 0.509455 |

**TABLE 2** Computer simulation parameters

| Parameter | Value |
|---|---|
| $f_c$ (GHz) | 5 |
| $d_{\text{Tx-Rx}}$ (m) | 55.73 |
| $d_{\text{Tx-RIS}}$ (m) | 35.51 |
| $d_{\text{RIS-Rx}}$ (m) | 20.22 |
| $h_{\text{Tx}}$ (m) | 10 |
| $h_{\text{RIS}}$ (m) | 4 |
| $h_{\text{Rx}}$ (m) | 1 |
| $W_0$ (dBm) | $-130$ |

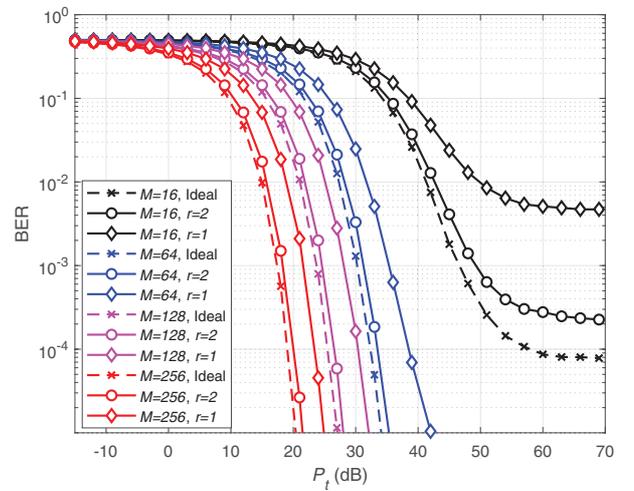

**FIGURE 6** Performance of discrete phase compared to ideal phase for varying $M$

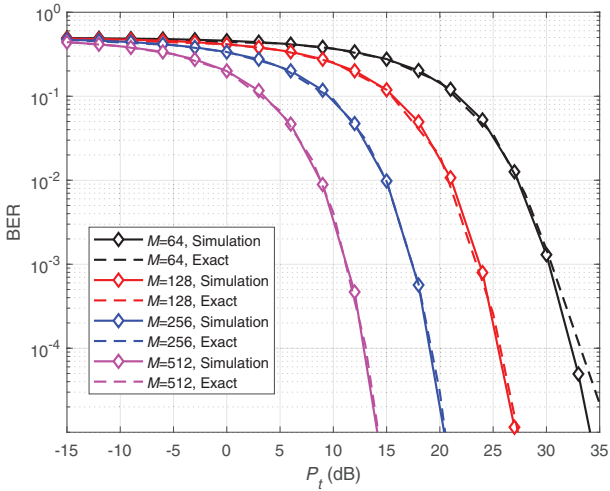

**FIGURE 5** Comparison of exact and simulated results for varying number of RIS elements $M$

Figure 6 compares the BER performances for ideal and discrete phase adjustments, where $r$ is the bit resolution of the discrete phase levels. For ideal phase adjustments, the phase term at the RIS is set exactly to eliminate the phase of the channel of the first tap. On the other hand, for discrete phase adjustments, the phase shifts at the RIS can only take certain values according to $r$ and it rounded to the closest phase term to maximize the first tap. For example, when $r = 2$, there are $2^r = 4$ possible phases to choose from to maximize the first tap. Since discrete phases may not completely align the phase terms of the first tap, the full potential of the first tap maximization and ISI elimination cannot be achieved. However, if the RIS is large enough, even 1 bit phases, $r = 1$, are enough to boost the desired signal terms. As it can be seen, the discrete case performs noticeably worse compared to the ideal case. For a large $M$, we do not observe an error floor, on the other hand, for a small number of RIS elements, such as $M = 16$, the RIS size is not sufficient to remove ISI and an irreducible error floor is observed. The parameter $M = 16$ is simply not sufficient number of RIS elements to provide the required performance similar to the case when using the central limit theorem, we would need a certain number of repetitions or

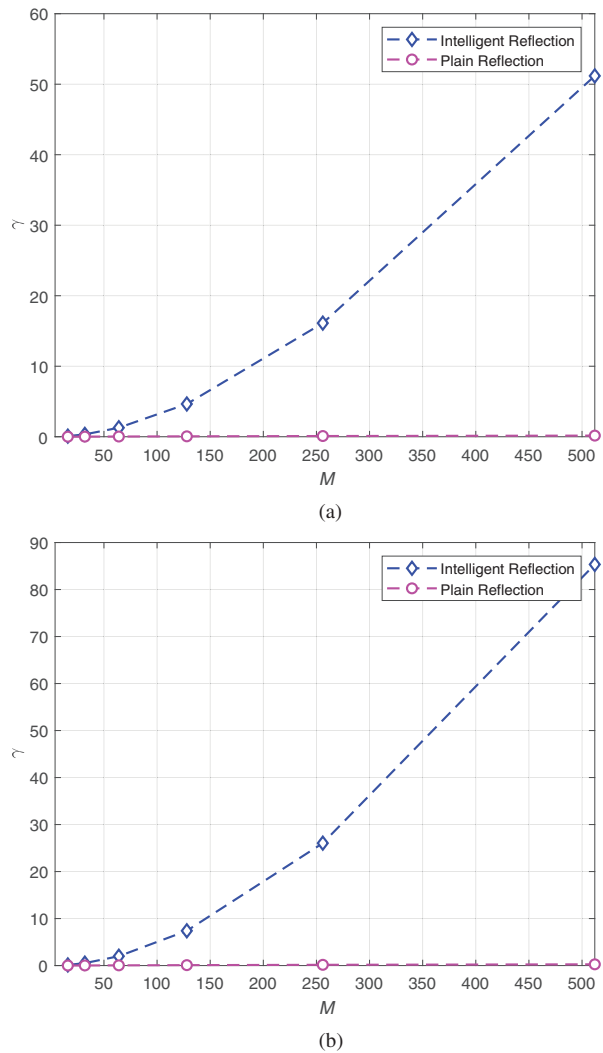

**FIGURE 7** (a) SINR comparison of the intelligent reflection and non-intelligent reflection for varying $M$ and (a) a uniform PDP and (b) an exponential PDP samples for it to satisfy. As the resolution is decreased from 2 to 1, the error floor increases drastically.

Figure 7 compares the SINR of the proposed over-the-air equalization scheme with a blind reflection from the RIS with no ISI elimination for $N_b = 4$, $P_t = 30$ dB and varying $M$ under uniform and exponential PDPs, respectively. By a plain and blind reflection, we consider a lossless reflection without any phase adjustment at the RIS elements, hence, not affecting the phases of the taps to eliminate the ISI. Figure 7a presents the performance improvement considering uniform PDP for the worst-case scenario. On the other hand, Figure 7b considers a more realistic exponential PDP. Comparing Figure 7a,b, it can be observed that the more realistic case, an exponential PDP, outperforms the worst-case uniform PDP scenario in terms of ISI minimization. This is because more power is allocated to the first tap in the exponential PDP case hence, exponentially reducing the power of the remaining ISI taps resulting in an increased SINR. It can also be concluded that the intelligent reflection significantly outperforms the plain reflection. This is because the ISI is made negligible for the intelligent reflection while in the plain reflection case, since equalization is not performed, ISI makes communication not possible which is seen by an almost zero SINR for any RIS size. When revisiting the BER curves and examining Figure 7 together, they complement each other where as $M$ increases, the SINR also exponentially increases thus, improving the BER.

## 6 | CONCLUSION

The main motivation of this study has been to provide a unique framework and perspective to RIS-based systems and virtual equalization for frequency-selective channels. It has been shown that an efficient over-the-air equalization concept, independent from the Tx and Rx, can be realized by utilizing the RIS. The RIS adjusts the phases of its elements according to the incoming signals to maximize the magnitude of the first tap, making the ISI terms negligible, hence, maximizing the SINR performance. A complete end-to-end system model and a probabilistic analysis of ISI elimination have been provided. In addition, a theoretical error probability analysis in conjunction with computer simulation results have shown the advantages of the proposed scheme. Future works may include an extension of this model to different scenarios, such as when a direct link exists between the Tx and Rx. Also, in the future this model may be analysed under different channel fading distributions, multiple RISs and for mobile nodes.

**FUNDING INFORMATION**
None.

**CONFLICT OF INTEREST**
The authors have declared no conflict of interest.

**DATA AVAILABILITY STATEMENT**
Data available on request from the authors

**ORCID**
*Emre Arslan*  https://orcid.org/0000-0002-0627-5051